\documentstyle[12pt]{article}
\begin{document}
\bibliographystyle{unsrt}

\newtheorem{theorem}{Theorem}
\newtheorem{lemma}{Lemma}
\newtheorem{defn}{Definition}

\newcommand{\be}{\begin{eqnarray}}
\newcommand{\ee}{\end{eqnarray}}
\newcommand{\ou}{\overline u}
\newcommand{\uu}{\underline u}
\newcommand{\hu}{\hat u}
\newcommand{\Pa}{Painlev\'e}

\title{A Local
Asymptotic Analysis of the First Discrete Painlev\'e Equation
as the Discrete Independent Variable Approaches Infinity}
\author{Nalini Joshi\\
        {\it School of Mathematics}\\
        {\it University of New South Wales}\\
        {\it Sydney NSW 2052}\\
	{\it Australia}\\
  	 {\tt N.Joshi@unsw.edu.au}
}
\maketitle
\begin{center}
Short Title: Asymptotics of dPI\\
AMS No's: 39A10, 41A60, 30E15\\
This paper is dedicated to Martin David Kruskal on the occasion
of his
seventieth birthday.
\end{center}
\begin{center}
{\bf Abstract}
\end{center}
The first discrete Painlev\'e equation (dPI), which appears in
a model of quantum gravity,
is an integrable
nonlinear nonautonomous difference
equation which yields the well known first Painlev\'e equation
(PI) in a continuum limit.
The
asymptotic study of its solutions as the discrete time-step
$n\to\infty$ is important both for physical application and
for checking the accuracy of its role as a numerical discretization
of PI. Here we show that the asymptotic analysis carried out by
Boutroux (1913) for PI as its independent variable approaches
infinity can also be achieved for dPI as its discrete independent
variable approaches the same limit.

\baselineskip24truept
\section[1]{Introduction}
Our aim is to study the equation
\begin{equation}
y_{n+1}+y_n+y_{n-1}={\alpha n+\beta\over y_n} +\gamma
\label{dPI}
\end{equation}
in the limit as $n\to\infty$. Eqn (\ref{dPI}) is known as
the discrete first Painlev\'e equation (dPI) because the scaling
limit $y_n=1+h\bigl(-2w(x)+c/2\bigr)$, $x=\sqrt hn-b/2-c^2/24$, with
$\alpha=h^{3/2}a$, $\beta=-\gamma+3+h^2b$,
$\gamma=6+hc$,
yields the classical
first Painlev\'e equation (PI)
\[
w''=6w^2+x,
\]
as $h\to 0$.

PI is the simplest of six  well known nonlinear
second-order ordinary differential equations (ODEs)
in the complex plane called the Painlev\'e equations.
Their characteristic property
that all
movable singularities of all solutions are poles is called
the Painlev\'e property.
Painlev\'e \cite{pp:acta}, Gambier \cite{gambier:acta}, and
Fuchs \cite{fuchs:annalen}
identified them (under some mild conditions) as the only
such equations with the Painlev\'e property whose general solutions
are new transcendental functions.

More recently, it was discovered by Ablowtiz, Ramani, and Segur
\cite{ars:jmp}
that such equations are closely related to
integrable partial differential equations \cite{as:siam,ac:cup}.
There is an arsenal of rich
techniques that
exhibit the special character of the \Pa\
equations as integrable equations. In particular,
Painlev\'e \cite{pp:bull} showed that they can be written
in terms ratios of entire functions,
Fuchs \cite{fuchs:annalen} and later Jimbo {\it et. al} \cite{jmms:physd}
(see also \cite{okamoto:86}) showed they
are isomonodromy conditions for associated linear ODEs,
and Joshi and Kruskal \cite{njmdk:conn1} showed that their connection problems
can
be solved directly.
Consequently, their solutions play a distinguished role as
nonlinear special functions. A natural development is to ask:
What are their
corresponding discrete versions?

The first insight into this question came from a discrete version of PI
derived in a model of quantum gravity \cite{bk,fik}. Subsequently,
a tour-de-force of work was accomplished by Grammaticos, Ramani and
collaborators (see \cite{grp}, \cite{rgh}
and references in \cite{gr:exeter}) to
answer
this question. They derived integrable discrete versions of the first to the
fifth Painlev\'e equations and identified a property, called the {\sl
singularity confinement} property, that appears to be a discrete version of the
Painlev\'e property.
Isomonodromy problems have been found
for these discrete \Pa\ equations. They appear, therefore, to be integrable and
deserve further study as rare examples of nontrivial integrable systems.
An
asymptotic description of their possible orbits as $n\to\infty$ is a natural
part of this study.

The limit we study, i.e. $n\to\infty$, is related to the continuum
limit. In such a limit, $n$ is typically multiplied by a small parameter
$h^{\mu}\to 0$, where $\mu=1/2$ for dPI to PI, while
$nh^{\mu}$ remains $O(1)$. Such a limit can only be achieved if $n$
is itself approaching infinity. In order to understand how the orbits of dPI
relate to the solutions of PI in a continuum limit, we need to study dPI in the
limit $n\to\infty$. This is crucial also for deciding whether or not dPI is
an accurate numerical discretization of PI.

It should be noted, however, that the limit $n\to\infty$ is not
equivalent to a continuum limit.
Continuum limits of dPI involve scaling limits of $y_n$ and the
parameters $\alpha$, $\beta$, $\gamma$, in addition to requiring $n\to\infty$.
Moreover, there exist limits of dPI other than the one described above to PI.
For example, an alternative continuum
limit to PII is also known. Nevertheless, for simplicity,
we restrict the comparison of our discrete results
to those known for PI.

We show that the asymptotic behaviours admitted by dPI as $n\to\infty$ are
qualitatively extremely similar to those found for PI by
Boutroux \cite{boutroux:I}. Boutroux
showed that in general the asymptotic behaviours of the (first four)
Painlev\'e transcendents near infinity in the plane of the independent variable
are given by elliptic functions. He studied the behaviour of
the transcendents along a line of approximate periods of such elliptic
functions and he showed that within certain sectors near infinity there
also exist special behaviours which are asymptotically free of poles.
His deductions were facilitated by a transformation of variables that made the
asymptotic behaviours more explicit. For PI, Boutroux's transformation
is
\be
w(x)&=&\sqrt xu(z)\nonumber\\
z&=&{4\over 5}x^{5/4}
\ee
under which PI becomes
\begin{equation}
u''=6u^2+1-{u'\over z}+{4\over 25}{u\over z^2},
\label{bPI}
\end{equation}
where the primes now denote differentiation with respect to $z$. (See
\cite{njmdk:conn1} for a deduction of Boutroux transformations based on maximal
dominant balances.) It is clear that on a path where $u$, $u'$, $u''$ are
bounded, and $u''$ is of order unity, the
leading-order behaviour of Eqn(\ref{bPI}) as $|z|\to\infty$ is given by
\[
u''\approx 6u^2+1,
\]
which is solved by Weierstrass elliptic functions. On the other hand, if
$u''\ll 1$, as $|z|\to\infty$, the leading-order behaviour is given by
\[
u^2\approx -\,{1\over 6},
\]
which gives rise to pole-free behaviours.

There exists a Boutroux transformation of dPI that also
facilitates its asymptotic study as $n\to\infty$. An analysis of
possible dominant balances of dPI as $n\to\infty$ shows that there is only
one maximal balance
\[
y_{n+1}+y_n+y_{n-1}\approx{\alpha n\over y_n},
\]
which is achieved when $y_n$ grows like $O(\sqrt n)$. Hence we are led
to
\[
y_n=\sqrt n u_n
\]
which transforms dPI to
\be
u_{n+1}+u_n+u_{n-1}&=&{\alpha\over u_n}+{\gamma\over\sqrt n}
             +{\beta\over nu_n}\nonumber\\
             &-&\left\{{\sqrt{n+1}-\sqrt n\over\sqrt n}\right\}u_{n+1}
             -\left\{{\sqrt{n-1}-\sqrt n\over\sqrt n}\right\}u_{n-1}.
\label{bdPI}
\ee
In Section 2, we show that this equation is solved to leading-order by
elliptic functions generally and by pole-free behaviours less
generally.

The elliptic-function-type behaviours are described by
an energy-like parameter $E$ and a phase-like parameter $\Phi$ which are
constant to leading-order but vary slowly as $n\to\infty$. The
evolution of $E$ over many periods of the leading-order elliptic function
can be studied via an averaging method. We carry out this study in
Section 3.

Although qualitatively we obtain very similar results to those found
by Boutroux, we point out here that there are differences.
In the special case $\gamma=0$, the main result
of Section 3 can be regarded as the discrete analogue of the
corresponding
result found by Boutroux for PI. However, in the case $\gamma\not=0$,
the slow evolution of $E$ has an additional component.
This additional complexity is due to the fact that our asymptotic
limit $n\to\infty$ is not exactly a continuum limit.
Moreover,
the limit $n\to\infty$ includes the case where $x=nh^{\mu}$ is
finite. Our analysis therefore includes regions where Boutroux's
asymptotic analysis may not be valid (he assumed $x\to\infty$).

What we demonstrate in
this paper is
that a comprehensive (formal) local asymptotic description of dPI can
be carried out as $n\to\infty$. Although our results are obtained
as $n\to\infty$, a similar leading-order analysis can be carried out
in neighbourhoods of any ordinary point $n=n_0$. This also yields
elliptic-function-type behaviours and pole-free behaviours.
Hence our analysis provides
futher evidence of the special
character of dPI as an integrable equation. Ramani, Grammaticos {\it et al}
have pointed out that dPI has the singularity confinement property
by a perturbation analysis carried out around its singularity
at $y=0$. Our asymptotic results verify
this pole-like singularity structure of the orbits to leading order
by showing that locally in patches near infinity,
the orbits are given by
meromorphic functions.

\section[2]{Local Asymptotic Results}
Here we analyse Equation (\ref{bdPI}) to find the locally valid asymptotic
behaviours of dPI in the limit $n\to\infty$. Our results are qualitatively
the same as those found by Boutroux for PI.

Suppose that $u_n$, $u_{n+1}$, $u_{n-1}$ are all $O(1)$.
Then we
have a maximal dominant balance of Eqn (\ref{bdPI}) given by
\begin{equation}
u_{n+1}+u_n+u_{n-1}\approx{\alpha\over u_n}.
\label{eft}
\end{equation}
Henceforth we adopt the notation
\[
u_{n+1}=:\overline u, u_n=u, u_{n-1}=:\underline u.
\]
Multiplying Eqn (\ref{eft}) by $u\ou$ and $u\uu$ respectively and
subtracting the two results, we get
\[
u\ou^2 + u^2\ou - u^2\uu - u\uu^2 \approx \alpha(\ou -\uu).
\]
Both sides of this equation are exact
differences and so we can integrate (actually sum)
to get
\begin{equation}
u\ou^2+u^2\ou - \alpha(\ou+u) = :E.
\label{Eeqn}
\end{equation}
Note that $E$ is
constant to leading-order
as $n\to\infty$, but varies slowly with $n$.
That is, $\overline E-E$ is not zero but
small (in fact, of order $O(1/\sqrt{n})$ for $n\gg 1$, see Eqn (14)).
We will refer to $E$ as an energy-like integral (or parameter) for dPI.
Consider the left side of Eqn(\ref{Eeqn}) as a polynomial in $x=u$, $y=\ou$.
The value of $E$ (to leading-order) defines level curves
of this polynomial which are
parametrized by elliptic functions (see \cite{AS:handbook}).

An alternative way to see that the solution involves elliptic functions
is to sum
the equation once more. Solve Eqn (\ref{Eeqn}) for $\ou$ and write the
result as
\[Q(u, E)= {1\over 2u}\left(\alpha -u^2\pm\sqrt{P(u, E)}
             \right)
\]
where
\[P(u, E):= (u^2+\alpha)^2+4Eu.\]
Then the result can be written as
\[
{u(\ou-\uu)\over \sqrt{P(u, E)}} = 1
\]
which upon integration gives
\[
\sum_{k=1}^n{u_k(u_{k+1}-u_k)\over \sqrt{P(u_k, E_k)}}= n-\Phi,
\]
where $\Phi$ is a constant.
The left side is a discrete analogue (Riemann sum)
of an elliptic integral of the
first kind.

The leading-order elliptic function has two periods.
The periods are (in general) complex
numbers given by the fact that the curve defined by Eqn
(\ref{Eeqn}) has genus two. In other words, $P$ has four branch
points (for generic values of $E$) and there exist two linearly
independent closed contours that enclose a pair of these. We will
denote the two periods (given by
the two contours $C_j$) as $\omega_j$,
$j=1, 2$.

In fact, the elliptic function is a discrete sampling of the
continuum (Jacobian) elliptic functions. So the definition
of $\omega_j$ requires slightly more explanation. We regard them
as being defined by the elliptic integral
\begin{equation}
\omega_j=\oint_{C_j}du\,{u\over \sqrt{P(u,E)}},
\end{equation}
where the path of integration for each given $E$
is a closed contour (as described
above) formed by interpolation through the
orbits of $u$ parametrized by the phase $\Phi$.

The elliptic-function-type behaviour $u$ of dPI, possesses these
periods only to leading-order. However, we nevertheless define
$\omega_j(E)$ in the same way and regard them as (implicit)
functions of $n$ through their dependence on $E$. We will also call
them (loosely) {\sl periods}.

There exist special values of $E$ for which the elliptic functions
degenerate to singly periodic (trigonometric) functions. To find
these values, rewrite $P$ as
\[P(u, E) = (u-\rho)^2(u^2+\sigma u + \tau).\]
Multiplying out the product on the right and equating coefficients
gives
\be
\sigma = 2\rho&,& \tau - 2\rho\sigma+\rho^2 = 2\alpha\\
\rho^2\sigma-2\rho\tau = 4E&,& \rho^2\tau = \alpha^2
\ee
Solving these shows that
\[\rho^2 = \alpha/3,\, - \alpha\]
and
\[E =  -{4\over 3}\alpha\sqrt{\alpha/3},\, 0\]
respectively. For these values of $E$, the elliptic integrals described
above degenerate to yield only a single period for $u$.

Now suppose that $u$, $\ou$, $\uu$ are still $O(1)$ but $\ou\approx u$,
$\uu\approx u$. Then the first integral (\ref{Eeqn}), although still
defined, describes an algebraic function (to leading order).
Since $u$ is constant to this order, where the constant is given by
\begin{equation}
3u \approx {\alpha\over u},
\label{s}
\end{equation}
we write $u=\sqrt{(\alpha/3)}+v$,
where $v\ll 1$.
Then Eqn (\ref{bdPI}) becomes
\be
\overline v+4v+\underline v &=& {\gamma\over\sqrt n}\bigl(1
	+v/\sqrt{(\alpha/3)}\bigr)\nonumber\\
& &+ O\Bigl(1/n, v\overline v, v^2,
v\underline v, \overline v/n, \underline v/n\Bigr)
\ee
We get the
asymptotic solution
\begin{equation}
u\approx\sqrt{\left(\alpha/3\right)}+ {\gamma\over 6\sqrt n}+ O(1/n).
\label{polefree}
\end{equation}
This solution obviously has an infinite algebraic asymptotic
series expansion and, therefore, has no poles for $n\gg 1$.
For that reason, we call such solutions {\sl pole-free} behaviours.

A straightforward calculation shows that a perturbation of a solution
$V$ with such behaviour, i.e. $u=V+\hat v$, gives
\[
\hat v\approx c_{\pm}\lambda_{\pm}^n
\]
where
\[
\lambda_{\pm} = -2 \pm \sqrt 3,
\]
and $c_\pm$ are arbitrary constants.
Obviously, $|\lambda_-|>1$ and $|\lambda_+|<1$. So to get a
consistent perturbation, we must have $c_-=0$.  In other words,
we get an exponentially small perturbation, multiplied by
$c_+$, which is hidden beyond
all orders of the algebraic expansion (\ref{polefree}) as
$n\to\infty$. Since the free parameter $c_+$ cannot be identified
uniquely, the pole-free behaviour (\ref{polefree}) cannot describe
a solution uniquely.

We chose the solution given by $\lambda_+$ above
under the assumption that $n$ is real, positive. However, in
general, dPI could be regarded as a mapping posed along a complex
line, i.e. $n$ could be complex. Therefore, the perturbation
$\hat v$ could be given by either $(\lambda_{\pm})^n$ or indeed
a combination of both (e.g. if $n$ is pure imaginary).
In other words, the behaviours given to the first few orders
by (\ref{polefree}) suffer from Stokes' phenomenon in the
complex $n$-plane.

There is one other dominant balance giving rise to pole-free
behaviours. In obtaining $V$ above, we assumed that
\[ u\approx c,\, \ou + \uu \approx 2c,\]
and then found the constant $c$ to be given by $\sqrt{(\alpha/3)}$.
An alternative consistent assumption is
\[u\approx c,\, \ou + \uu \approx -2c.\]
In this case, we find that
\begin{equation}
u\approx (\pm 1)^n\sqrt{-\alpha}+ O(1/\sqrt n).
\label{polefree2}
\end{equation}
Again the solutions with this leading order behaviour suffer
from Stokes' phenomenon. However, now the Stokes' lines
are orthogonal to those of the pole-free behaviour given
by $V$. Writing $u=W+\hat w$, where $W$ is given to leading
order by Eqn(\ref{polefree2}), we get
\[\hat w \approx b_{\pm}(\pm i)^n,\]
where $b_\pm$ are arbitrary constants. So along
purely real directions in the $n$-plane, the perturbation
$\hat w$ is approximately oscillatory, whereas along
purely imaginary directions, one of $b_\pm$ must
be zero and $\hat w$ is hidden beyond all orders of the
divergent expansion for $W$.

Along anti-Stokes' lines, i.e. directions along which
the perturbations $\hat v$ or $\hat w$ are approximately
oscillatory, the perturbations are not simply linear
combinations of the exponentials found above. As for
PI (see \cite{njmdk:connPI,njmdk:conn1}),
they are more accurately represented by (discrete)
Fourier series which yield an expansion composed of increasing
powers of such exponentials multiplied by algebraic prefactors.
We omit the details here for simplicity.

It can be shown
that the elliptic-function-type behaviours and the pole-free
behaviours are connected in the space of solutions. The latter
can be shown to be
degenerate limits of the former, attained when
\hfill\break $E\to-(4\alpha/3)\sqrt{\alpha/3}$ or $E\to 0$.

Although the number of pole-free behaviours of PI (or PII) differs from
that for dPI
(PI has two, PII has three),
these results reflect qualitatively
exactly the asymptotic results found by Boutroux.

\section[3]{Averaging}
In this section, we analyse the change of the energy-like parameter
$E$ (describing the elliptic-function-type behaviours)
over a period $\omega_j$ of the leading-order elliptic function
by using an averaging method.

$E$ fluctuates with $n$. This is clear from the definition
(\ref{Eeqn}) which gives
\be
E- \underline E&=&(\ou -\uu)\Biggl\{{\gamma u\over \sqrt n}
		+{\beta\over n}
	 	 -\biggl(\sqrt{(1+1/n)}-1\biggr)\ou u\nonumber\\
	 	& &\quad -\biggl(\sqrt{(1-1/n)}-1\biggr) u \uu\Biggr\}
\label{Eexpn}
\ee

Suppose we are given initial values at some point $\Phi$, $|\Phi|\gg
1$, which define a solution as follows:
\[
u(\Phi)=1,\quad u(\Phi + 1) = p\, (\not=0).
\]
Then we have
\[
E(\Phi)=p^2+p-\alpha(p+1).
\]
Assume the periods $\omega_j$ to be those given at $\Phi$ by this
value of $E$.

To carry out an averaging method, we assume that there is a
smooth, slowly varying function
interpolated between the successive points on the orbit defined
by $E$ and $\Phi$ (or the initial values above).
We wish to study the slow variation of $E$ from
one period of the leading-order elliptic function to the next.
Let $u$ be
\[u=U+ s,\]
where $U$ represents the slowly-varying part of $u$ and $s\ll U$
represents
its fast fluctuations (as $|n|\to\infty$).
The initial values are now
\be
U(\Phi)&=& 1, \quad U(\Phi+1)=p,\nonumber\\
s(\Phi)&=& 0, \quad s(\Phi+1)=0.
\ee
Then Eqn (\ref{bdPI}) gives
\be
\overline U + U + \underline U&=&{\alpha\over U}\\
\overline{s} + s + \underline{s}&=&\alpha
     \Biggl({1\over U+s}-{1\over U}\Biggr)
     +{\beta\over n(U+s)}+{\gamma\over\sqrt n}\nonumber\\
    & &-\Biggl(\sqrt{1+1/n}-1\Biggr)\bigl(\overline U+\overline{s}\bigr)
       -\Biggl(\sqrt{1-1/n}-1\Biggr)\bigl(\underline
U+\underline{s}\bigr)\nonumber\\
\ee
Note that $U$ is an elliptic function (by arguments presented in
Section 2).

Consider the first integral given by Eqn(\ref{Eeqn}).
Expanding the solution $u$ as $U+s$ and keeping only terms
of order $s$ or $1/\sqrt n$ gives
\begin{equation}
\overline{s}(2\overline UU+U^2-\alpha)+s({\overline U}^2
     +2\overline UU -\alpha)=\gamma{\overline UU\over\sqrt n}.
\label{hatuint}
\end{equation}
This equation can be integrated once more by using the following
observations. The defining equation for $U$ gives
\be
\overline UU - U\underline U&=& \overline UU +U(\overline U
              +U-\alpha/U)\\
                  &=&2\overline UU+U^2-\alpha\\
-\overline{\overline U}\,\overline U + \overline UU &=& \overline U(\overline
U+U-\alpha/\overline U) + \overline UU\\
                  &=& {\overline U}^2+2\overline UU -\alpha
\ee
Letting
\[F:= \overline UU - U\underline U,\]
we have
\[\overline F = \overline{\overline U}\,\overline U - \overline UU.\]
Hence Eqn(\ref{hatuint}) becomes
\[
\overline{s}F - s \overline F = \gamma{\overline UU\over\sqrt
n}.
\]
There is an integrating factor for this equation. After division by
$\overline F
F$, it becomes
\[
\left(\,\overline{\left({s\over F}\right)}-{s\over F}\,\right)=
\gamma{\overline UU\over\overline F F\sqrt
n}.
\]
Hence we can integrate (actually sum) to get $s$ as
\[
s_n = \gamma F_n\sum_{k=\Phi}^{n-1} {U_{k+1}U_{k}\over
F_{k+1}F_{k}\sqrt{k}},
\]
where the sum is understood to be zero in the case $n=\Phi$.

Now consider Eqn(\ref{Eexpn}) for $E$. Note that the first term on
the right gives
\[
E\bigl(\Phi + \omega\bigr) - E\bigl(\Phi\bigr)
  = \gamma {\overline UU\over\sqrt n}\Biggm|^{\Phi+\omega}_{\Phi}
        +O\bigl(1/\Phi, s/\sqrt{\Phi}\bigr),
\]
to leading order. However, the contribution of this term
is actually smaller than $O(1/\sqrt{n})$ because
$U$ is periodic and
\[
{1\over\sqrt{\Phi +
\omega}}-{1\over\sqrt{\Phi}}=O\left(1/\Phi^{3/2}\right).
\]
To get the nonzero slow change of $E$, therefore,
we need to consider terms of size $O(1/n, s/\sqrt n)$.

At these orders, we get
\be
E\bigl(\Phi + \omega\bigr) - E\bigl(\Phi\bigr)
 &\approx& \gamma {\overline U s+U\overline{s}\over\sqrt n}
          \Biggm|^{\Phi+\omega}_{\Phi}\\
 & &\quad +{\beta (\overline U+ U)\over
n}\Biggm|^{\Phi+\omega}_{\Phi}\\
  & &\quad  -{1\over 2\Phi}\sum_{k=\Phi+1}^{\Phi+\omega}
               U_k(U_{k+1}-U_{k-1})^2
\ee
By periodicity and integration by parts, the second ratio on the right
can be shown to be zero to leading order.
Hence we have
\be
E\bigl(\Phi + \omega\bigr)&-&E\bigl(\Phi\bigr)\nonumber\\
   &\approx&\quad{\gamma^2\over\Phi} \Biggl\{U_{n+1}F_n\sum_{k=\Phi}^{n-1}
					{U_{k+1}U_{k}\over
					F_{k+1}F_{k}}
                                   +U_nF_{n+1}\sum_{k=\Phi}^{n}
                                        {U_{k+1}U_{k}\over
                                        F_{k+1}F_{k}}
                        \Biggr\}\Biggm|^{\Phi+\omega}_{\Phi}\nonumber\\
  & &\qquad  -{1\over 2\Phi}\sum_{\Phi+1}^{\Phi+\omega}
               U_k(U_{k+1}-U_{k-1})^2\nonumber\\
   &\approx&\quad{\gamma^2\over\Phi}
\biggl(U_{\Phi+1}F_{\Phi}+U_{\Phi}F_{\Phi+1}\biggr)
                  \sum_{k=\Phi}^{\Phi+\omega-1}
                                        {U_{k+1}U_{k}\over
                                        F_{k+1}F_{k}}
                                                \nonumber\\
  & &\qquad  -{1\over 2\Phi}\sum_{\Phi+1}^{\Phi+\omega}
               U_k(U_{k+1}-U_{k-1})^2
\label{Eevol}
\ee
To simplify this expression, we use the notation $P(u, E)$
defined in Section 2. Note that
\[F= U(\overline U-\underline U) =  \sqrt{P(U, E)}.\]
Therefore, we have
\[{\overline UU\over\overline F F} = {\overline UU\over\sqrt{P(U, E)P(\overline
U,
\overline E)}}.
\]
Hence Eqn(\ref{Eevol}) becomes
\be
E\bigl(\Phi + \omega\bigr)&-& E\bigl(\Phi\bigr)\nonumber\\
   &\approx& {\gamma^2\over\Phi}
       \biggl(U_{\Phi+1}F_{\Phi}+U_{\Phi}F_{\Phi+1}\biggr)
               \sum_{k=\Phi}^{\Phi+\omega-1}
					{U_{k+1}U_{k}\over
                                     \sqrt{P(U_{k+1},E)P(U_{k},E)}}\nonumber\\
  & &\quad  -{1\over 2\Phi}\sum_{\Phi+1}^{\Phi+\omega}
               {P(U_k,E)\over U_k}
\label{Esimple}
\ee

In the case $\gamma=0$, this result yields
\begin{equation}
E\bigl(\Phi + \omega\bigr)-E\bigl(\Phi\bigr)\approx-{1\over
2\Phi}\sum_{\Phi+1}^{\Phi+\omega}
               {P(U_k,E)\over U_k}.
\label{zerogamma}
\end{equation}
This is the discrete analogue of the result found by Boutroux.
To recognize this, note that
\[{P(U_k,E)\over U_k} = (U_{k+1}-U_{k-1})\sqrt{P(U_k,E)},\]
and so
the sum on
the right of Eqn (\ref{zerogamma})
is a discrete analogue of the elliptic integral
\[\tilde\omega_j=\oint_{C_j}\,du\,\sqrt{P(u,E)}.\]
(Note that the derivative of this integral with respective to $E$
is the previously described elliptic integral $\omega_j$ and
that in making continuum analogies we have ignored numeric
factors.)
This is what Boutroux found (see
page 319 of \cite{boutroux:I}) for the evolution of $E$ (called
$D$ in his paper). He showed by using this result that
along a chain of increasing points $\Phi_j$ defined by
$\Phi_0=\Phi$, $E_0=E(\Phi)$, $\Phi_j=\Phi_{j-1}+\omega(E_{j-1})$,
$E_j=E(\Phi_j)$, $E$ must be bounded and approach a value for
which the leading order elliptic function degenerates to
a trigonometric function.

The generic case $\gamma\not=0$
differs markedly from this (see the remarks made near the end of
Section 1) and will
form the subject of future explorations.

\section[]{Acknowledgements}
It is a pleasure to thank the
Isaac Newton Institute, where this study was first started, and the Australian
Research Council for their support.

\bibliography{refs}
\end{document}